\documentclass[twocolumn]{aastex701}
\usepackage{subcaption}
\usepackage{amsmath}

\begin{document}

\title{Tracing black hole and galaxy growth across environments since cosmic noon}

\author[orcid=0009-0000-6612-0599]{Emma Jane Weller}
\affiliation{Department of Astronomy, Yale University, New Haven, CT 06511, USA}
\email[show]{emma.weller@yale.edu}

\author[0000-0002-5554-8896]{Priyamvada Natarajan}
\affiliation{Department of Astronomy, Yale University, New Haven, CT 06511, USA}
\affiliation{Department of Physics, Yale University, New Haven, CT 06511, USA}
\affiliation{Black Hole Initiative, Harvard University, Cambridge, MA 02138, USA}
\email{}

\author[0000-0001-9947-6911]{Colin J. Burke}
\affiliation{Department of Astronomy, Yale University, New Haven, CT 06511, USA}
\email{}

\author[0000-0002-7941-1149]{Shashank Dattathri}
\affiliation{Department of Astronomy, Yale University, New Haven, CT 06511, USA}
\email{}

\begin{abstract} 
The distribution of systems in the black hole (BH) mass--stellar mass ($M_\mathrm{BH}-M_\star$) plane encodes both the integrated growth of galaxies and their central black holes, and the processes that shape their evolution. Using the ASTRID and TNG300 cosmological simulations, we track massive BHs from cosmic noon ($z=2$) to $z=0.5$. TNG300 repositions BHs to the centers of their host galaxies at every time step, while ASTRID instead advances them under the resolved gravitational forces plus a subgrid dynamical friction model, allowing BHs to wander off-center when orbital decay is inefficient. We follow the BHs rather than their original host galaxies, thereby capturing central BHs, BHs in satellites, and wandering BHs. We find that central BHs in both simulations evolve along a tight, nearly redshift-invariant $M_\mathrm{BH}-M_\star$ relation that is broadly consistent with local empirical constraints. Departures from this relation trace distinct evolutionary channels in which mergers play a key role. Major galaxy mergers drive BH-BH coalescence that dominates the growth of the most massive central BHs. These BHs subsequently quench their hosts through active galactic nucleus (AGN) kinetic feedback. Minor mergers tidally strip satellites to lower $M_\star$ at nearly fixed $M_\mathrm{BH}$, producing weakly accreting, overmassive central BHs in gas-poor systems. In ASTRID, satellite accretion and inefficient dynamical friction generate wandering BHs that are undermassive relative to their new hosts and experience minimal accretion or merger-driven growth. A BH's location in the $M_\mathrm{BH}-M_\star$ and specific BH accretion rate--specific star formation rate planes is therefore a fossil record of its dynamical, accretion, merger, and feedback history.
\end{abstract}


\section{Introduction} \label{sec:intro}

It is well understood that the masses of central black holes (BHs) in the local Universe are correlated with properties of the stellar components of their host galaxies, suggesting a close connection between central BH growth and galaxy assembly. However, debate remains over the nature of this coevolution and the relative importance of BH seeding versus feedback-regulated growth over cosmic time. Furthermore, the time dependence of the $M_\mathrm{BH} - M_\star$ relation is not yet well understood. In this work, we trace the evolution of populations of BHs and their host galaxies using two cosmological simulations, ASTRID and the TNG300 volume of the IllustrisTNG suite. In addition to central BHs, we follow wandering BHs, whose scaling relations with their host galaxies have not been well studied.

\subsection{BH--galaxy scaling relations} \label{sec:scaling}

Empirical scaling relations connecting the mass of a central BH to properties of the host bulge, including its stellar mass, velocity dispersion, and luminosity, have been extensively measured at low redshift (e.g. \citealt{Magorrian+98, Ferrarese_Merritt_2000, Gebhardt+2000, Haring_Rix_2004, Kormendy_Ho_2013, McConnell_Ma_2013}). Mass measurements of high-redshift galaxies, now increasingly common with James Webb Space Telescope (JWST) data, show that their central BHs tend to be overmassive compared to the local $M_\mathrm{BH} - M_\star$ relation (e.g. \citealt{Harikane+2023, Maiolino+2024, Yue+2024, Kocevski+2025}). However, the evolution of the relation over time is still poorly constrained. Some studies conclude that JWST's overmassive BHs can largely be explained by selection bias and are consistent with the local relation (e.g. \citealt{Li+2025, Sun+2025_z6, Sun+2025_z4}), while others suggest intrinsic evolution of the relation (e.g. \citealt{Pacucci+2023, Maiolino+2024, Yue+2024}).

The implications of the $M_\mathrm{BH} - M_\star$ relation for BH-galaxy coevolution have been widely investigated using models and simulations. These studies show that the relation is shaped by many processes. A key factor is active galactic nucleus (AGN) feedback, through which massive BHs heat and expel gas from the galaxy, quenching star formation and BH accretion (e.g. \citealt{Silk_Rees_1998, DiMatteo+2005, Hopkins+2006, Kulier+2015}). The prescriptions for supernova feedback, gas accretion, and dynamics also impact the relation (e.g. 
\citealt{Somerville_Dave_2015, Habouzit+2017, Habouzit+2021, Dattathri+25}). Both heavy and light seeding models can reproduce local scaling relations, but heavy seeding models predict BHs that are initially overmassive relative to local relations, while light seeding models predict initially undermassive BHs (e.g. \citealt{Volonteri_Natarajan_2009, Volonteri2010, PN2014, Bhowmick+2025_seeds}). The JWST observations of very overmassive BHs at high redshift (see Sec. \ref{sec:scaling}) may imply the existence of heavy seeds (e.g. \citealt{Andika+2024, Pacucci_Loeb_2024}). A notable example is UHZ1, an X-ray luminous AGN at $z \approx 10.1$ that is highly consistent with theoretical predictions for a heavy seed formed via direct collapse of gas \citep{Goulding+2023, Bogdan+2024, Natarajan+2024}.

\subsection{Wandering BHs} \label{sec:wanderers}

Most studies of massive BHs, including those described in the previous section, focus on central BHs (also referred to as centrals in this work). However, simulations like ASTRID (see Sec. \ref{sec:astrid}) and ROMULUS \citep{Tremmel+2017}, which do not pin BHs to the local potential minimum, reveal a largely undetected population of massive BHs wandering off-center in their host galaxies. The in-depth study of these BHs (often called wanderers) is relatively recent (see, e.g., \citealt{Bellovary+2010, Gonzalez_Guzman_2018, Tremmel+2018, Greene+2021, Ricarte+2021_unveiling, Ricarte+2021_origins, Seepaul+2022, Weller+2022, Weller+2023, DiMatteo+2023}). Simulations \citep{Bellovary+2019} and observations \citep{Reines+2020} suggest that a significant fraction of massive BHs in dwarf galaxies are off-nuclear; in simulations, this is mainly caused by galaxy-galaxy mergers \citep{Bellovary+2021}. In many cases, when a satellite galaxy hosting a massive BH merges with a larger galaxy, the satellite is tidally disrupted, and its BH is left wandering on a wide orbit. Dynamical friction is ineffective at sinking these BHs toward the galactic center, so many remain wandering for several Gyr, resulting in hundreds of wanderers per halo \citep{Tremmel+2018, Ricarte+2021_origins, DiMatteo+2023}.

Wandering BHs typically orbit in low-density regions -- often well outside the galactic center and with large orbital inclinations relative to the galactic plane -- so they experience minimal accretion, which makes them challenging to detect \citep{Ricarte+2021_origins, DiMatteo+2023, Weller+2023}. However, if they retain a stellar component, they may dominate the rates of tidal disruption events at masses $\lesssim 10^7 \, \mathrm{M}_\odot$. These events can produce flares of detectable electromagnetic and gravitational waves \citep{Fragione+2018, Ricarte+2021_unveiling, Pfister+2022}. 

\subsection{This work} \label{sec:structure}

As discussed in Sec. \ref{sec:scaling}, many studies have investigated the $M_\mathrm{BH} - M_\star$ relation in cosmological simulations, including IllustrisTNG and -- to a lesser extent -- ASTRID (see, e.g., \citealt{Li+2020, Terrazas+2020, Habouzit+2021, Habouzit+2022, Haidar+2022, Dattathri+25}). However, most of these studies focus on the overall scaling at different redshifts and mass ranges. In this work we take a different approach, selecting a sample of (mostly) central BHs at $z = 2$ (cosmic noon) and tracking them to $z = 0.5$ (a period of 5.3 Gyr). By this point, observations and theoretical models suggest that the bulk of BH and galaxy growth is complete \citep{Shankar+2013, Madau_Dickinson_2014}, and the $M_\mathrm{BH} - M_\star$ relations in ASTRID and TNG300 are largely established and close to local empirical relations (\citealt{Habouzit+2021, Ni+2025}; see also Fig. \ref{fig:Median_Mbh-Mstar_lines}). 

ASTRID and TNG300 have different prescriptions for BH seeding, dynamics, and feedback. We examine how these differences impact BH growth, star formation, and the resulting $M_\mathrm{BH} - M_\star$ relations. Because we follow the BHs rather than the galaxies, a subset of the central BHs in ASTRID merge into more massive galaxies and end up as wanderers. While most of the work on wandering BHs has investigated their dynamics and observational signatures (see Sec. \ref{sec:wanderers}), we focus instead on their growth history over cosmic time and their evolution through different host environments.

The paper is structured as follows. Sec. \ref{sec:simulations} provides an overview of ASTRID and TNG300, and Sec. \ref{sec:selection} describes our BH sample selection. In Sec. \ref{sec:evolution} we investigate the evolution of these BHs relative to their host galaxies, using plots of BH mass vs. stellar mass (Sec. \ref{sec:Mbh-Mstar}) and specific BH accretion rate vs. specific star formation rate (Sec. \ref{sec:sBHAR-sSFR}). In Sec. \ref{sec:observations}, we compare our results to the local empirical $M_\mathrm{BH} - M_\star$ relation. Finally, in Sec. \ref{sec:conclusion}, we summarize our findings and conclude the paper.

\section{Simulations} \label{sec:simulations}

\subsection{ASTRID} \label{sec:astrid}

ASTRID \citep{Bird+2022, Ni+2022} is a cosmological simulation run with MP-Gadget \citep{Feng+2018}, a smoothed particle hydrodynamics code that uses the TreePM algorithm for gravitational interactions. ASTRID started at $z=99$ and recently reached $z=0$ \citep{Zhou+2026}. It contains $5500^3$ dark matter particles and an initially equal number of gas particles in a box with side length $250 \, h^{-1} \, \mathrm{cMpc}$. The dark matter and initial gas particle masses are $6.74 \times 10^6 \, h^{-1} \, \mathrm{M}_\odot$ and $1.27 \times 10^6 \, h^{-1} \, \mathrm{M}_\odot$, respectively. The gravitational softening length is $1.5 \, h^{-1} \, \mathrm{ckpc}$. Cosmological parameters follow \cite{Planck2020}. Halos and subhalos are identified using the friends-of-friends (FOF, \citealt{Davis+1985}) and SUBFIND \citep{Springel+2001} algorithms, respectively. Below, we summarize the implementation of BH seeding, accretion, dynamics, merging, and feedback (for more details, see \citealt{Ni+2022, Ni+2025}). Subgrid models for additional astrophysical processes are implemented as described in \cite{Bird+2022, Ni+2022}.

BHs are seeded in halos with $M_\mathrm{tot} > 5 \times 10^9 \, h^{-1} \, \mathrm{M}_\odot$ and $M_\star > 2 \times 10^6 \, h^{-1} \, \mathrm{M}_\odot$ that do not yet contain a BH. The BH seed mass is drawn from a power-law probability distribution with index $n=-1$, ranging from $3 \times 10^4 \, h^{-1} \, \mathrm{M}_\odot$ to $3 \times 10^5 \, h^{-1} \, \mathrm{M}_\odot$. The densest gas particle in the halo is converted into a BH particle with the position and velocity of the parent gas particle. The accretion of gas onto the BH is estimated using a Bondi-Hoyle-Lyttleton-like prescription \citep{DiMatteo+2005}. Unlike many cosmological simulations, ASTRID does not pin BHs to the centers of their hosts. Instead, it uses a subgrid dynamical friction model for BH dynamics \citep{Tremmel+2015, Chen+2022}, allowing for a population of wandering BHs, as described in Sec. \ref{sec:wanderers}. Two BHs are marked as having merged if their separation is less than twice the gravitational softening length, and their kinetic energy has been dissipated such that they are gravitationally bound.

For a given BH, AGN feedback operates in either a high-accretion quasar mode or a low-accretion radio mode (sometimes called jet mode). At $z > 2.3$, BHs operate only in the high-accretion mode.  At $z \leq 2.3$, a BH enters the low-accretion mode when:
\begin{equation} \label{eqn:fdbk1}
    f_\mathrm{Edd} < \chi_\mathrm{thr},
\end{equation}
where $f_\mathrm{Edd} = \dot{M}_\mathrm{BH} / \dot{M}_\mathrm{Edd}$ is the BH's Eddington ratio (accretion rate divided by the Eddington rate) and $\chi_\mathrm{thr}$ is a BH mass-dependent threshold given by:
\begin{equation} \label{eqn:fdbk2}
    \chi_\mathrm{thr} = \min{\left[ 0.002 \left( \frac{M_\mathrm{BH}}{M_\mathrm{crit}} \right)^2, \, \chi_\mathrm{cap} \right]},
\end{equation} 
where $M_\mathrm{crit} = 5 \times 10^8 \, h^{-1} \, \mathrm{M}_\odot$ and $\chi_\mathrm{cap} = 0.05$. This means that the low-accretion mode typically activates only for very massive BHs.

The feedback energy in the high-accretion mode is given by:
\begin{equation} \label{eqn:fdbk3}
    \Delta \dot{E}_\mathrm{high} = \epsilon_\mathrm{f,th} \epsilon_\mathrm{r} \dot{M}_\mathrm{BH} c^2,
\end{equation}
where $\epsilon_\mathrm{f,th} = 0.05$ is the coupling efficiency and $\epsilon_\mathrm{r} = 0.1$ is the radiative efficiency. The energy is injected into the surrounding gas as thermal energy, so we refer to this mode as AGN thermal feedback. 

The feedback energy in the low-accretion mode is given by:
\begin{equation} \label{eqn:fdbk4}
    \Delta \dot{E}_\mathrm{low} = \epsilon_\mathrm{f,kin} \dot{M}_\mathrm{BH} c^2,
\end{equation}
where the coupling efficiency $\epsilon_\mathrm{f,kin}$ increases with the local gas density and has a maximum value of $0.05$. The feedback energy is injected into the surrounding gas as bursts of momentum, so this mode is called AGN kinetic feedback. It is less susceptible to cooling losses than thermal feedback, and therefore more efficient at quenching star formation in massive galaxies \citep{Weinberger+2017, Weinberger+2018, Nelson+2018, Weller+2025}.

\subsection{TNG300} \label{sec:tng}

IllustrisTNG \citep{Marinacci+2018, Naiman+2018, Nelson+2018, Pillepich+2018_results, Springel+2018} is a suite of cosmological simulations run from $z=127$ to $z=0$ using AREPO \citep{Springel2010}, a moving-mesh code that, like MP-Gadget, uses TreePM to solve gravitational interactions. In this work, we use TNG300-1 (hereafter TNG), the highest-resolution run of the largest simulation volume, which most closely matches the ASTRID volume. TNG contains $2500^3$ dark matter particles and an initially equal number of gas cells in a box with a side length of $205 \, h^{-1} \, \mathrm{cMpc}$. The dark matter particle mass and the target baryon mass are $4.0 \times 10^7 \, h^{-1} \, \mathrm{M}_\odot$ and $7.6 \times 10^6 \, h^{-1} \, \mathrm{M}_\odot$, respectively. Dark matter and stellar particles have a $z=0$ gravitational softening length of $1.0 \, h^{-1} \, \mathrm{ckpc}$. TNG uses cosmological parameters from \cite{Planck2016}. As in ASTRID, halos and subhalos are identified using the FOF and SUBFIND algorithms, respectively. The TNG galaxy formation model is described in \cite{Pillepich+2018_methods}. Below, we summarize the BH modeling (for more details, see \citealt{Weinberger+2017}).

In contrast to ASTRID, TNG uses a single, relatively high BH seed mass of $8 \times 10^5 \, h^{-1} \, \mathrm{M}_\odot$. A BH is seeded whenever a halo that does not yet contain a BH reaches $M_\mathrm{tot} > 5 \times 10^{10} \, h^{-1} \, \mathrm{M}_\odot$. BH accretion is estimated using an Eddington-limited Bondi accretion rate. At every global integration time step, each BH is repositioned to the local potential minimum (the location of minimum gravitational potential within a region containing the equivalent of 1000 mass resolution elements). BHs merge once they are within their ``feedback radius'' (the scale on which AGN thermal feedback energy is applied to the surroundings). Because of the BH repositioning, mergers occur rapidly \citep{Vogelsberger+2013, Springel+2018}, and there is no wandering BH population.

As in ASTRID, AGN feedback operates in either a high-accretion thermal mode or a low-accretion kinetic mode. These modes follow the same form as equations \ref{eqn:fdbk1} - \ref{eqn:fdbk4}, but with different parameter values: $M_\mathrm{crit} = 10^8 \, \mathrm{M}_\odot$, $\chi_\mathrm{cap} = 0.1$, $\epsilon_\mathrm{r} = 0.2$, $\epsilon_\mathrm{f,th} = 0.1$, and $\epsilon_\mathrm{f,kin}$ has a maximum value of $0.2$. Unlike in ASTRID, there is no redshift limit on the low-accretion kinetic mode.

Because of the well-understood modeling differences between ASTRID and TNG, comparing results derived from the two simulations allows us to better understand the impacts of AGN kinetic feedback and BH repositioning.

\section{Selection of BH sample} \label{sec:selection}

We began by identifying galaxies (corresponding to SUBFIND subhalos) at $z=2$ with $M_\star > 10^9 \, \mathrm{M}_\odot$. This stellar mass limit ensures that our galaxies are resolved with $\gtrsim 1000$ (ASTRID) or $\gtrsim 100$ (TNG) stellar particles.\footnote{In ASTRID, an average gas particle produces 4 stellar particles, so the stellar particle mass is $\sim 10^6 \, \mathrm{M}_\odot$ \citep{Bird+2022}. In TNG, the average initial stellar particle mass is roughly equal to the target baryon mass, which is $\sim 10^7 \, \mathrm{M}_\odot$ \citep{Pillepich+2018_results}.} Throughout this work, all galaxy properties (namely, stellar mass and star formation rate) are measured within twice the stellar half-mass radius.

We also required the galaxies to contain at least one BH with mass $> 10^6 \, \mathrm{M}_\odot$ (ASTRID) or $> 10^{6.5} \, \mathrm{M}_\odot$ (TNG). This BH mass limit excludes BHs that are still close to their seed masses and keeps our analysis outside of the regime where the $M_\mathrm{BH} - M_\star$ relation is artificially modified by seeding effects \citep{Habouzit+2021}. Due to TNG's higher BH seed mass and rapid BH mergers (caused by repositioning), its BH mass function is more skewed toward high masses than ASTRID's.

We classified each of the galaxies that met our criteria as ``primary,'' ``satellite,'' or ``isolated.'' In a halo with multiple galaxies, we labeled the one with the most bound particles (generally equivalent to the central galaxy in a galaxy group) as the primary, and the others as satellites. If a halo had only one galaxy, we called the galaxy isolated. We randomly selected 5000 primary and 5000 satellite galaxies in each simulation at $z=2$.

As noted previously, this work focuses on following the BHs rather than their host galaxies. In each galaxy, we identified the most massive BH, which is usually near the center of the galaxy (but not always -- in ASTRID, $5\%$ of our BHs are already outside their host's stellar half-mass radius when they are selected at $z=2$). We then tracked the BH to $z=1$ and $z=0.5$ using BH merger catalogs \citep{Blecha+2016, Kelley+2017, Chen+2022_mergers, Ni+2022}. However, many BHs (mainly in TNG) were not traceable via this method due to incompleteness in the merger catalogs. Additionally, if multiple BHs in our sample merged into the same BH or the same host galaxy, we kept only the first BH in our list. We dropped BHs that ended up in an isolated galaxy or a SUBFIND subhalo with zero stellar mass. In total, the number of BHs in our sample decreased from 10000 at $z=2$ to 9309 (ASTRID) and 4146 (TNG) at $z=0.5$. Despite the significant decrease in TNG, the surviving sample is sufficiently large that our results remain robust. We note, however, that the mass gained from mergers by TNG BHs (shown in Fig. \ref{fig:Merger-gain_Mbh-Mstar}) may be underestimated due to missing merger events.

At each of the three snapshots ($z = 2, 1, 0.5$), we identified properties of the BHs and their host galaxies and halos. While the halos are not a focus of this paper, their masses are shown in Fig. \ref{fig:Mhalo_hist} of Appendix \ref{app:halo_masses} to provide a comprehensive picture of mass assembly.

To place our selected BHs in context, we randomly chose 50000 galaxies with nonzero stellar mass at $z=0.5$ in each simulation. In each galaxy with at least one BH (26759 galaxies in ASTRID and 23591 in TNG), we identified the most massive BH (usually the central), and used these BHs to compute the simulation's ``typical'' $z=0.5$ $M_\mathrm{BH} - M_\star$ relation. This relation is shown for comparison in all of our $M_\mathrm{BH} - M_\star$ plots.

\section{Evolution of BH populations} \label{sec:evolution}

\subsection{BH mass--stellar mass plane} \label{sec:Mbh-Mstar}

\begin{figure*}
    \centering
    \includegraphics[width=2\columnwidth]{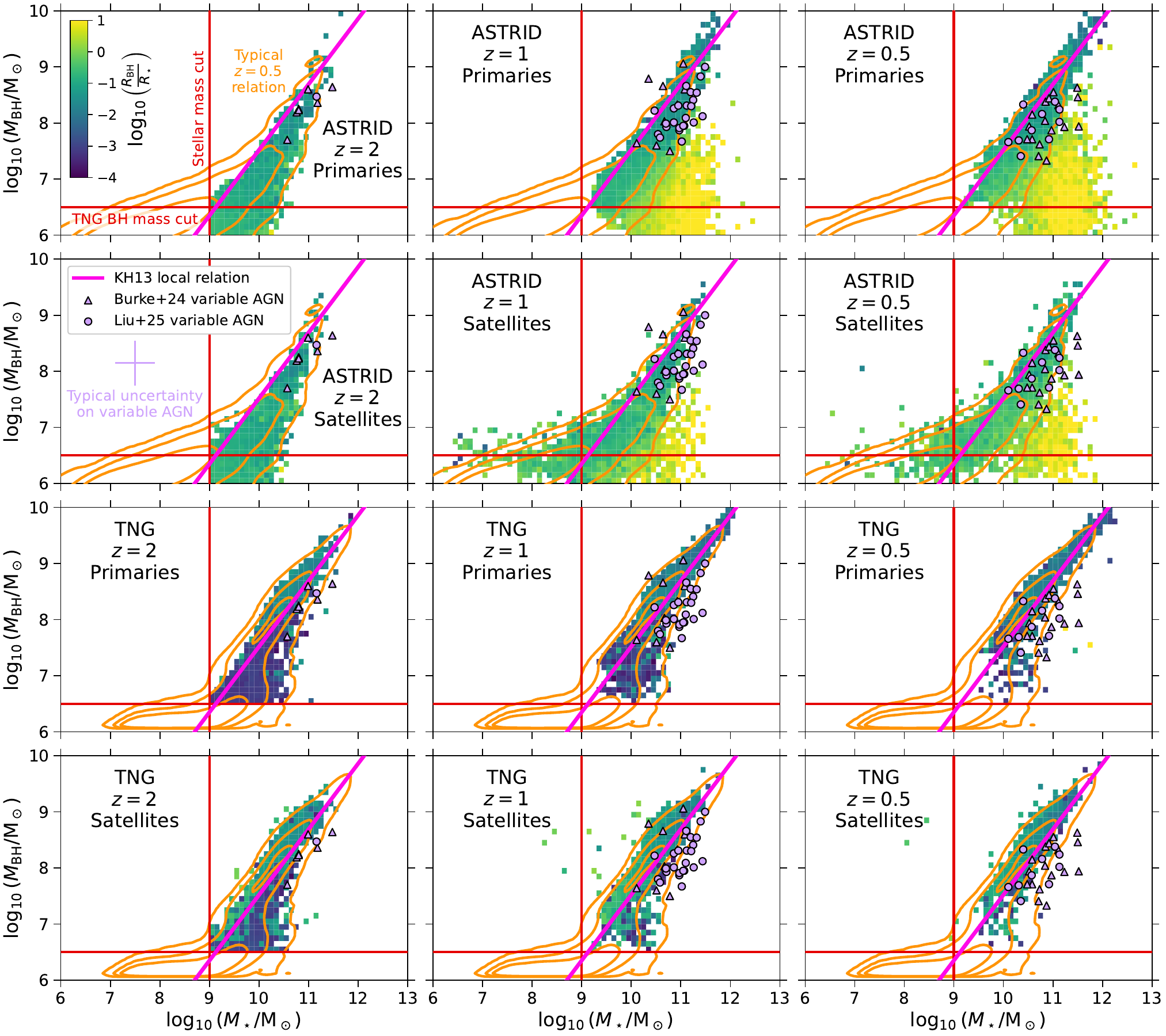}
    \caption{Binned $M_\mathrm{BH} - M_\star$ maps for our BH sample, with each bin colored by the median value of $\log_{10}{(R_\mathrm{BH}/R_\star)}$, where $R_\mathrm{BH}$ is a BH’s distance from the center of its host galaxy and $R_\star$ is the galaxy's stellar half-mass radius. Panels are separated by simulation, redshift, and host galaxy type. We also plot $1\sigma$, $2\sigma$, and $3\sigma$ contours of the typical $z=0.5$ $M_\mathrm{BH} - M_\star$ relation in the simulations. For comparison, we add the local ($z \sim 0$) empirical relation from KH13 and variable AGN from \cite{Burke+2024} and \cite{Liu+2025} at $z \sim 2, 1, 0.5$. The clump of points that forms to the right of the typical $M_\mathrm{BH} - M_\star$ relation in ASTRID represents a population of wandering BHs. The central BHs are broadly consistent with these observational results, with TNG more closely matching the KH13 relation and ASTRID better covering the region occupied by variable AGN.}
    \label{fig:Rbh_Mbh-Mstar}
\end{figure*}

\begin{figure*}[!t]
    \centering
    \includegraphics[width=1.33\columnwidth]{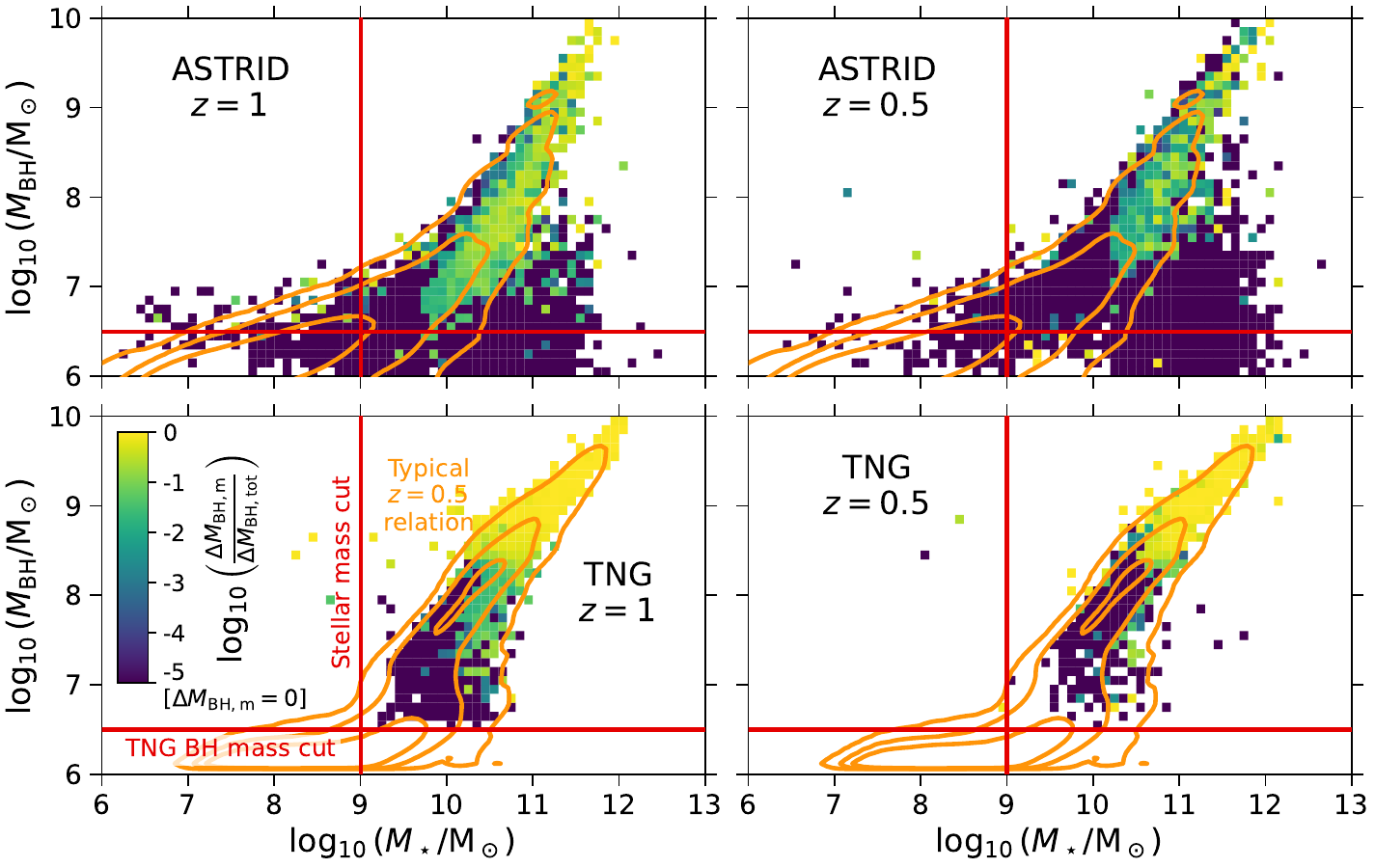}
    \caption{Binned $M_\mathrm{BH} - M_\star$ maps for our BH sample at $z=1$ and $z=0.5$, with each bin colored by the median value of $\log_{10}{(\Delta M_\mathrm{BH,m} / \Delta M_\mathrm{BH,tot})}$, where $\Delta M_\mathrm{BH,tot}$ is a BH's total mass gain since the previous snapshot ($z=2$ or $z=1$), and $\Delta M_\mathrm{BH,m}$ is the amount that came from mergers. BHs with $\Delta M_\mathrm{BH,m} = 0$ are set to $\Delta M_\mathrm{BH,m} / \Delta M_\mathrm{BH,tot} = 10^{-5}$ for visualization. In TNG, the higher-mass BHs grow almost entirely via mergers, and the lower-mass BHs grow primarily via accretion. In ASTRID, the more massive BHs also experience mergers, but not as much as in TNG. The wanderers and the BHs in stripped satellites undergo very few mergers. In general, there is more merger-driven growth between $z=2$ and $z=1$ than between $z=1$ and $z=0.5$.}
    \label{fig:Merger-gain_Mbh-Mstar}
\end{figure*}

\begin{figure*}[!t]
    \centering
    \includegraphics[width=2\columnwidth]{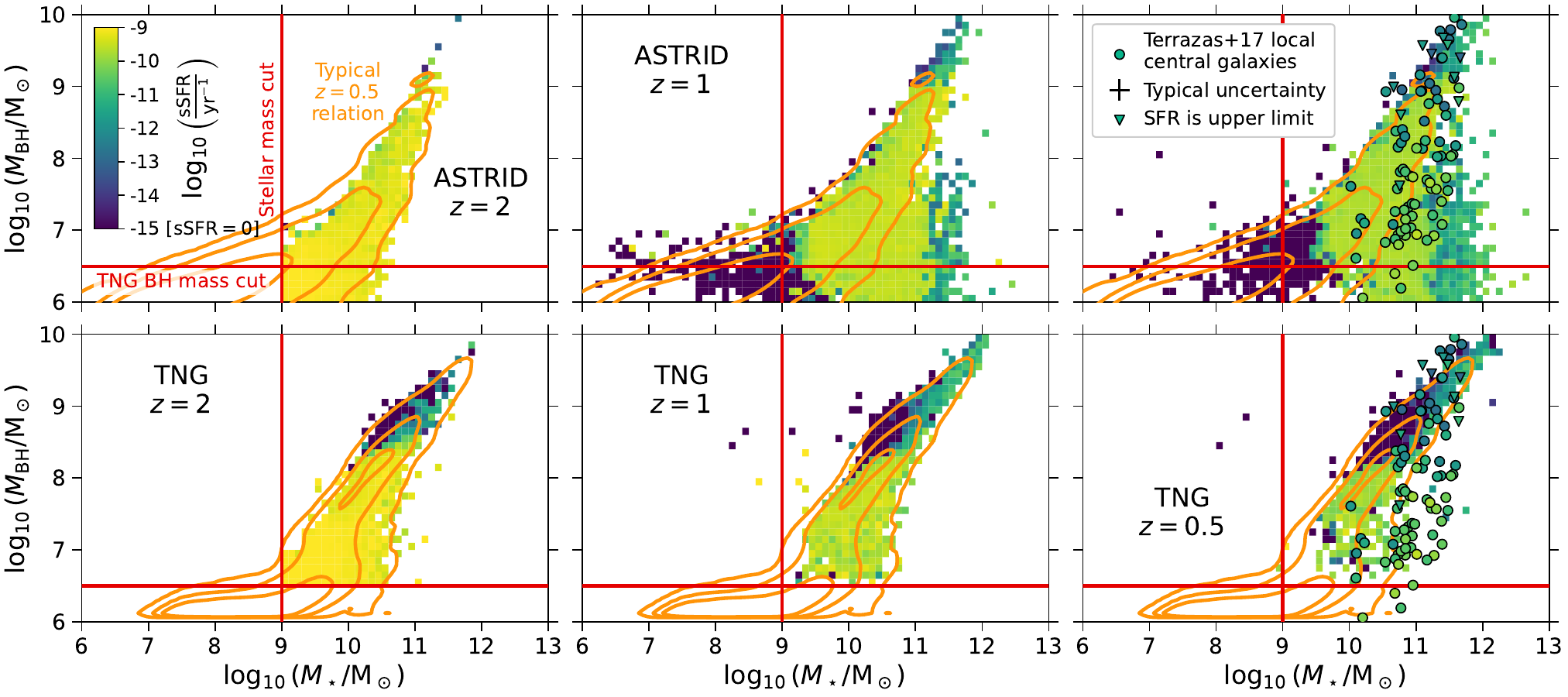}
    \caption{Binned $M_\mathrm{BH} - M_\star$ maps for our BH sample, with each bin colored by the median value of $\log_{10}{(\mathrm{sSFR}/\mathrm{yr}^{-1})}$, where $\mathrm{sSFR} = \mathrm{SFR}/M_\star$ is the specific star formation rate of a BH's host galaxy. Galaxies with $\mathrm{sSFR} = 0$ are set to $\mathrm{sSFR} = 10^{-15} \, \mathrm{yr}^{-1}$ for visualization. Panels are separated by simulation and redshift. The most massive galaxies and the stripped satellites have minimal star formation due to loss of cold gas caused by AGN feedback and tidal stripping, respectively. The star-forming galaxies have an approximately uniform sSFR that decreases with time. We add for comparison local ($z \sim 0$) observational data from \cite{Terrazas+2017}, and find that the simulations, especially TNG, exhibit narrower scatter and a sharper drop in sSFR at the high-mass end.}
    \label{fig:sSFR_Mbh-Mstar}
\end{figure*}

In Figs. \ref{fig:Rbh_Mbh-Mstar}, \ref{fig:Merger-gain_Mbh-Mstar}, and \ref{fig:sSFR_Mbh-Mstar}, we plot binned maps of BH mass vs. host galaxy stellar mass for our sample of BHs, with bins colored by a variety of properties. The color scale is normalized to the $0.1-99.9$ or $1-99$ percentile range of the plotted property across both simulations and all three snapshots, with the limits rounded to the nearest integers. The figure panels are separated by simulation and redshift, and, in Fig. \ref{fig:Rbh_Mbh-Mstar}, by host galaxy type (primary or satellite). We also show $1\sigma$, $2\sigma$, and $3\sigma$ contours of the typical $z=0.5$ $M_\mathrm{BH} - M_\star$ relation described in Sec. \ref{sec:selection}. We used Gaussian kernel density estimation to approximate the probability density function, and calculated contours enclosing $68.3\%$, $95.4\%$, and $99.7\%$ of the distribution.

In Fig. \ref{fig:Rbh_Mbh-Mstar}, we color each bin by the median BH galactocentric distance, $R_\mathrm{BH}$ (defined as the distance between a BH and the particle with the minimum potential energy in the BH's host galaxy), normalized by the host galaxy's stellar half-mass radius, $R_\star$. Most of the \textit{central} BHs (low $R_\mathrm{BH}/R_\star$) follow a roughly linear $M_\mathrm{BH} - M_\star$ relation, as also seen in Fig. \ref{fig:Median_Mbh-Mstar_lines}. The relation is similar between ASTRID and TNG, suggesting that the slope and normalization are only weakly impacted by the differences in AGN feedback prescriptions between the two simulations, in agreement with \cite{Angles-Alcazar+2015, Angles-Alcazar+2017}. The scatter in the linear relation is relatively small and approximately constant over time, consistent with e.g. \cite{Yang+2019, Shankar+2020, Habouzit+2021, Zou+2024}. However, this conflicts with the observed variance in local AGN, and with studies suggesting that BH mergers can decrease the scatter over time (e.g. \citealt{Peng_2007, Hirschmann+2010, Jahnke_Maccio_2011, Tanaka+2026}).

The points below the initial stellar mass cut represent satellite galaxies that have been tidally stripped. TNG only has a few such galaxies, likely because once galaxies are close enough for tidal stripping to occur, the BH from the satellite often gets classified as part of the primary and is quickly repositioned to the primary's center. The points extending from the low-mass end of the linear relation toward higher stellar mass have large galactocentric distances, so they represent a population of wandering BHs. The wanderers show no clear trend between $M_\mathrm{BH}$ and $M_\star$. In TNG, there is no such population since BHs are pinned to the centers of their hosts. In general, the central BHs in TNG are even closer to the centers of their hosts than the central BHs in ASTRID. TNG does have a few BHs with larger galactocentric distances; these are likely cases where a central BH in a smaller galaxy gets classified as part of a nearby larger galaxy but has not yet been repositioned to the center of the larger galaxy.

There is no sharp boundary between central and wandering BHs in $R_\mathrm{BH}$ or in $M_\mathrm{BH} - M_\star$ space. The 3$\sigma$ contour of the typical $M_\mathrm{BH} - M_\star$ relation roughly marks the left edge of the clump of wandering BHs with clearly distinct properties. This boundary approximately corresponds to a distance cut of $\log_{10}{(R_\mathrm{BH} / R_\star) \sim 0.5}$. Throughout this section and the following Sec. \ref{sec:sBHAR-sSFR}, we refer to the BHs inside the 3$\sigma$ contour as centrals and those to the right of the boundary as wanderers. In Sec. \ref{sec:observations}, we use a more conservative distance cut to select a robust set of central BHs.

For comparison to our simulation results, we show in Fig. \ref{fig:Rbh_Mbh-Mstar} the local empirical relation from \cite{Kormendy_Ho_2013} (hereafter KH13). TNG aligns more closely with this relation than ASTRID; for more on this, see Sec. \ref{sec:observations}. We note that the KH13 relation uses the stellar mass of the bulge, but that is not available in the simulations. Instead, we used the stellar mass within twice the galaxy's stellar half-mass radius. We also plot observational $M_\mathrm{BH} - M_\star$ data for optically variable AGN in the redshift ranges $z=2.5-1.5$, $1.25-0.75$, and $0.75-0.25$, from the Hyper Suprime-Cam Subaru Strategic Program \citep{Burke+2024} and the Dark Energy Survey Supernova Program \citep{Liu+2025}. These studies cover redshift, BH mass, and stellar mass ranges similar to our work, but we note that they include only AGN, while our sample includes both active and inactive BHs. In general, their AGN fall within our typical $M_\mathrm{BH} - M_\star$ relation in ASTRID, but extend slightly below the typical TNG relation.

There are few observational counterparts to the BHs in stripped satellites and the wandering BHs. This is unsurprising, given that these BHs typically exist in gas-poor environments and therefore have very low accretion rates, making them difficult to detect. However, they are beginning to be observed. For example, dynamical analyses of ultracompact dwarf galaxies reveal very overmassive BHs in what are likely the tidally stripped nuclei of initially larger satellites \citep{Seth+2014, Ahn+2017, Ahn+2018, Afanasiev+2018, Taylor+2025}. Wandering BH candidates have been identified via offset tidal disruption events \citep{Lin+2018, Jin+2025, Yao+2025, Guolo_2026, Stein+2026}, and in cases where they \textit{do} experience significant accretion, via hyper-luminous X-ray sources \citep{Farrell+2009, Farrell+2012, Barrows+2019, Tranin+2024} and offset AGN \citep{Barrows+2025, Barrows_Comerford_2025}.

\begin{figure*}[!t]
    \centering
    \includegraphics[width=2.1\columnwidth]{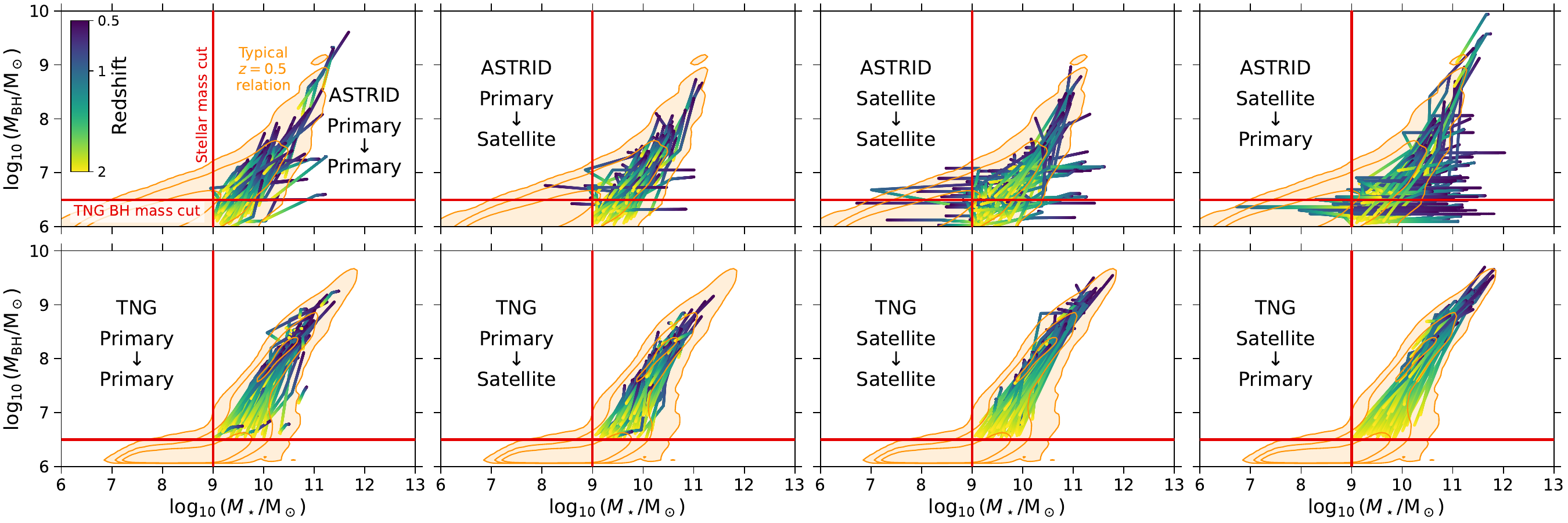}
    \caption{$M_\mathrm{BH}$ vs. $M_\star$ evolution of subsets of our BH sample. In each simulation, we randomly select 100 BHs that begin in a primary galaxy at $z=2$ and remain in a primary at $z=0.5$, 100 that begin in a primary and end up in a satellite, and so on. We locate each BH at $z=2,1,0.5$ and plot its track between these three snapshots. Most central BHs move upward along the typical $M_\mathrm{BH} - M_\star$ relation. Interactions and eventual mergers between galaxies cause tidal stripping (seen here as horizontal jumps to the left) and leave BHs from small galaxies wandering in more massive galaxies (horizontal jumps to the right). The most massive central BHs in primaries typically arise from mergers with BHs from satellites.}
    \label{fig:Mbh-Mstar_tracks}
\end{figure*}

In Fig. \ref{fig:Mbh-Mstar_tracks}, we plot the evolution of subsets of our sample from $z=2$ to $z=1$ to $z=0.5$ in the $M_\mathrm{BH} - M_\star$ plane. Fig. \ref{fig:Primary_Mbh-Mstar} in Appendix \ref{app:extra_plots} provides an alternative way of visualizing this evolution. In both ASTRID and TNG, central BHs in primary galaxies generally evolve upward along the simulation's $M_\mathrm{BH} - M_\star$ relation. The majority have been in primaries since $z = 2$. However, the highest-mass BHs mostly trace back to relatively massive satellites at $z=2$. These represent cases where major mergers cause multiple BHs to coalesce and form a very massive central BH.

The central BHs in satellites come from both primary and satellite galaxies. Most move upward along the $M_\mathrm{BH} - M_\star$ relation, but a subset of satellites in ASTRID shows large stellar mass drops with minimal changes in BH mass, indicating tidal stripping. The wandering BH population in ASTRID arises largely from galaxy mergers, as indicated by large increases in stellar mass with minimal BH growth, sometimes following a stellar mass drop corresponding to tidal stripping. Most of the wanderers originate as central BHs in satellites and merge into primaries, but satellites also host a wandering population, and a small number of wanderers originate in primaries. For a schematic of the different evolutionary tracks described above, see Fig. \ref{fig:Schematic}.

Fig. \ref{fig:Merger-gain_Mbh-Mstar} shows the fraction of BH growth that comes from mergers. For each BH, we calculated the mass gain from mergers between two snapshots as the total mass of all BHs that merged with the studied BH during this time period. We note that BH mergers in the simulations neglect mass loss due to gravitational radiation, but these losses are expected to be small relative to the total mass gains. We see from the figure that the most massive central BHs in TNG grow almost entirely via mergers (the mean fraction of BH growth that comes from mergers is $89\%$ at $1<z<2$ and $82\%$ at $0.5<z<1$ for $M_\mathrm{BH} > 10^9 \, \mathrm{M}_\odot$), while the less massive centrals grow primarily via accretion. In ASTRID, the wandering BHs and the central BHs in stripped satellites undergo few mergers (the mean fraction of growth from mergers is $9\%$ at $1<z<2$ and $4\%$ at $0.5<z<1$ for wanderers with $R_\mathrm{BH} / R_\star > 10^{0.5}$; $5\%$ and $3\%$ for BHs in stripped satellites with $M_\star < 10^9 \, \mathrm{M}_\odot$). The more massive central BHs experience mergers, but not as much as in TNG, which is expected given that TNG's BH repositioning procedure causes BH mergers to occur more rapidly. The merger-driven growth of the most massive centrals is consistent with our conclusions from Fig. \ref{fig:Mbh-Mstar_tracks}. The higher relative importance of mergers in TNG for the most massive centrals has little impact on the shape of the $M_\mathrm{BH} - M_\star$ relation, which is similar between the two simulations, as noted earlier in this section. In both simulations, more of the BHs undergo mergers between $z=2$ and $z=1$ than between $z=1$ and $z=0.5$ (the mean fraction of growth from mergers is $13\%$ at $1<z<2$ and $11\%$ at $0.5<z<1$ for ASTRID; $17\%$ and $13\%$ for TNG). For more on the buildup of BH mass via different growth channels, see \cite{Porras-Valverde+26}.

Fig. \ref{fig:sSFR_Mbh-Mstar} shows the stellar mass growth of our host galaxies. The star-forming galaxies have similar specific star formation rates (star formation rate per unit stellar mass, $\mathrm{sSFR} = \mathrm{SFR}/M_\star$) at a given redshift, and the typical sSFR decreases from $z=2$ to $z=0.5$. Most of the stripped galaxies have no star formation, likely due to the removal of their cold gas reservoirs. The most massive galaxies also have low sSFR; this can be attributed to AGN kinetic feedback, which activates for very massive BHs and quenches galaxies by heating and expelling cold gas. This effect is more extreme and extends to lower masses in TNG, likely because TNG has no redshift limit and a lower critical BH mass than ASTRID for the onset of AGN kinetic feedback (see Sec. \ref{sec:simulations}). For related plots showing the BH accretion rates, see Fig. \ref{fig:fEdd_Mbh-Mstar} in Appendix \ref{app:extra_plots}.

In the $z=0.5$ panels of Fig. \ref{fig:sSFR_Mbh-Mstar}, we overlay observational data for local central galaxies from \cite{Terrazas+2017}. Compared to these observations, we see that the simulations, especially TNG, display narrower scatter in the $M_\mathrm{BH} - M_\star$ relation (for central BHs) and a more abrupt drop in sSFR with increasing stellar and BH mass. We note that our simulation results are for primary and satellite galaxies at $z=0.5$, while the observations are from central (primary) galaxies at $z \sim 0$. However, our findings are consistent with \cite{Terrazas+2020}, who compared the \cite{Terrazas+2017} observational results to $z=0$ primary galaxies in IllustrisTNG.

\subsection{sBHAR--sSFR plane} \label{sec:sBHAR-sSFR}

\begin{figure*}[!t]
    \centering
    \includegraphics[width=1.8\columnwidth]{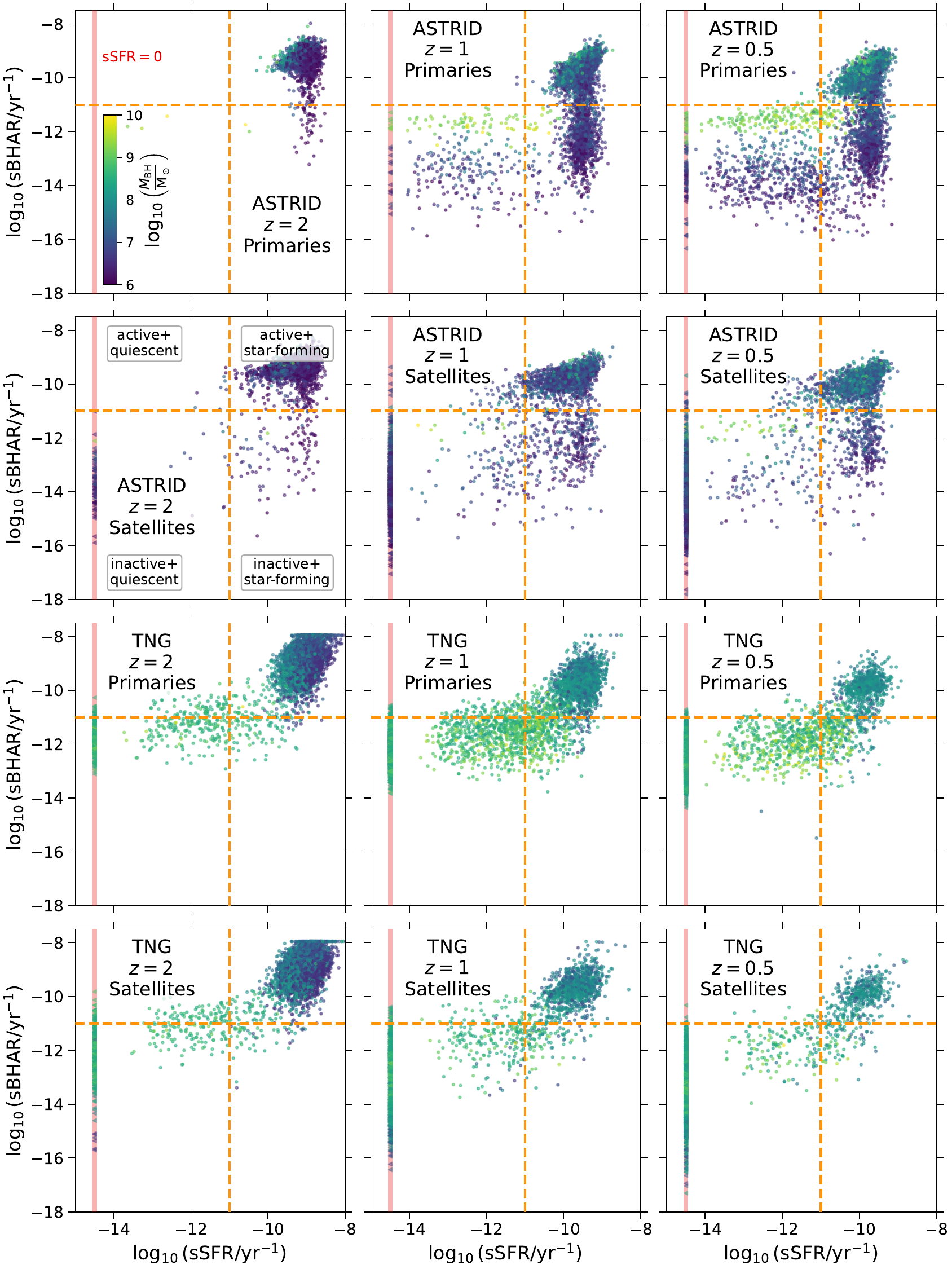}
    \caption{$\rm sBHAR$ vs. $\rm sSFR$ scatterplots for our BH sample, with each point colored according to the BH's mass. Panels are separated by simulation, redshift, and host galaxy type. The orange dashed lines roughly divide the plane into four quadrants: active+quiescent, active+star-forming, inactive+star-forming, and inactive+quiescent. Galaxies with $\mathrm{sSFR} = 0$ are marked with triangles and set to $\mathrm{sSFR} = 10^{-14.5} \, \mathrm{yr}^{-1}$ for visualization. Most points begin in the active+star-forming quadrant at $z=2$. The most massive BHs form a band that mostly lies in the inactive+quiescent quadrant, while less massive central BHs tend to stay in the active+star-forming quadrant.}
    \label{fig:Mbh_sBHAR-sSFR_quadrants}
\end{figure*}

\begin{figure*}[!t]
    \centering
    \includegraphics[width=2.1\columnwidth]{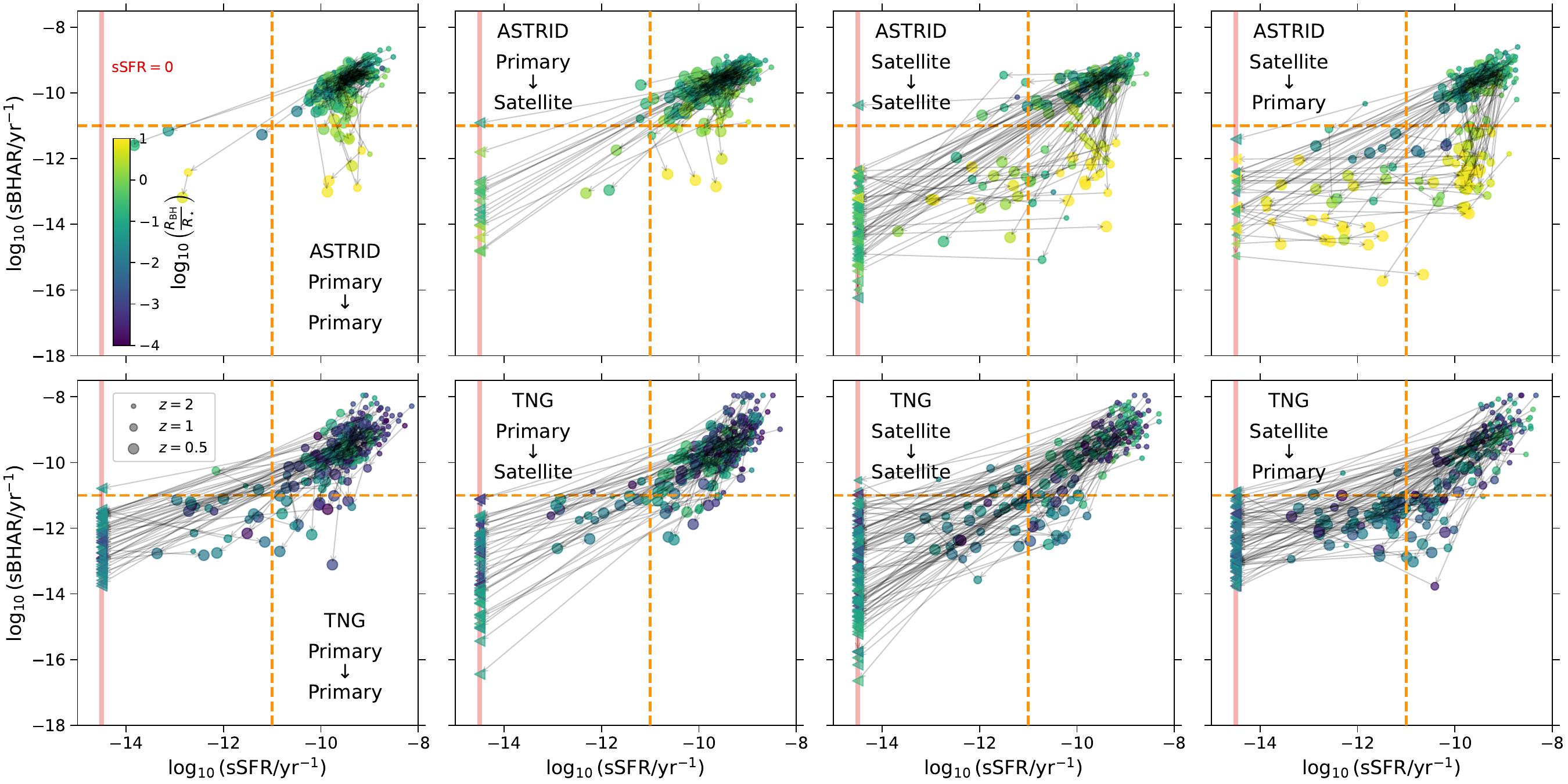}
    \caption{Evolution in the $\mathrm{sBHAR} - \mathrm{sSFR}$ plane for the same subsets of our BH sample used in Fig. \ref{fig:Mbh-Mstar_tracks}, with each point colored according to the BH's galactocentric distance. As in Fig. \ref{fig:Mbh_sBHAR-sSFR_quadrants}, the orange dashed lines divide the plane into four quadrants, and galaxies with $\mathrm{sSFR} = 0$ are marked with triangles and set to $\mathrm{sSFR} = 10^{-14.5} \, \mathrm{yr}^{-1}$ for visualization. Wandering BHs have low accretion rates but exist in both quiescent and star-forming galaxies, so they end up in the lower two quadrants. Tidally stripped galaxies usually have no star formation, so they form a vertical line at $\mathrm{sSFR} = 0$ in the inactive+quiescent quadrant.}
    \label{fig:Rbh_sBHAR-sSFR_tracks}
\end{figure*}

In order to study the growth histories of the various BH populations and their host galaxies, we compare the specific BH accretion rate (accretion rate per unit BH mass, $\mathrm{sBHAR} = \dot{M}_\mathrm{BH} / M_\mathrm{BH}$) to the host galaxy sSFR. As shown in \cite{Dattathri+25}, the $\mathrm{sBHAR} - \mathrm{sSFR}$ plane can be roughly divided into four quadrants: active+quiescent, active+star-forming, inactive+star-forming, and inactive+quiescent. Active and inactive here refer to the accretion state of the BH, while quiescent and star-forming refer to properties of the stellar component. We demarcate these quadrants with lines at $\mathrm{sBHAR} = 10^{-11} \, \mathrm{yr}^{-1}$ and $\mathrm{sSFR} = 10^{-11} \, \mathrm{yr}^{-1}$, as seen in Figs. \ref{fig:Mbh_sBHAR-sSFR_quadrants} and \ref{fig:Rbh_sBHAR-sSFR_tracks}. \cite{Dattathri+25} studied only primary galaxies and selected the most massive BH in each (which usually, but not always, corresponds to the central BH). They found that for central BHs, the location in the $\mathrm{sBHAR} - \mathrm{sSFR}$ plane encodes details about the host galaxy's gas content and quenching process. Systems in the active+star-forming quadrant generally correspond to low-to-intermediate mass, gas-rich galaxies, whereas systems in the inactive+quiescent quadrant correspond to high-mass galaxies that have been quenched by AGN kinetic feedback. Systems in the active+quiescent quadrant correspond to galaxies with slightly overmassive BHs that are undergoing a compaction-like quenching event. Finally, systems in the inactive+star-forming quadrant mostly correspond to off-center BHs that are unable to accrete efficiently despite residing in gas-rich galaxies or to galaxies undergoing gradual ``inside-out" quenching driven by AGN thermal feedback.

In contrast to \cite{Dattathri+25}, our analysis includes populations of wanderers and BHs in satellites. In Figs. \ref{fig:Mbh_sBHAR-sSFR_quadrants} and \ref{fig:Rbh_sBHAR-sSFR_tracks}, we color points in the $\mathrm{sBHAR} - \mathrm{sSFR}$ plane by BH mass and galactocentric distance, respectively. The color scales are normalized as described in Sec. \ref{sec:Mbh-Mstar}. In Fig. \ref{fig:Mbh_sBHAR-sSFR_quadrants} we show all of the BHs in our sample, while in Fig. \ref{fig:Rbh_sBHAR-sSFR_tracks} we trace the evolution of the same subsets of our sample used in Fig. \ref{fig:Mbh-Mstar_tracks}.

At $z=2$, most of the points in ASTRID are in the active+star-forming quadrant, reflecting the gas-rich states of the host galaxies. As time progresses, the most massive BHs tend to move into the inactive+quiescent quadrant, forming a horizontal band just below the sBHAR dividing line. As explained in Sec. \ref{sec:Mbh-Mstar}, these BHs have efficiently quenched their host galaxies through AGN kinetic feedback. The wandering BHs have low accretion rates due to their low gas density environments, but their hosts cover a range of sSFR values, so they end up in the inactive+quiescent and inactive+star-forming quadrants. As seen in Fig. \ref{fig:sSFR_Mbh-Mstar}, most of the stripped satellites have no star formation, so they fall into the inactive+quiescent quadrant, along the $\mathrm{sSFR} = 0$ line. The remaining BHs -- less massive central BHs along ASTRID's $M_\mathrm{BH}-M_\star$ relation -- typically stay in the active+star-forming quadrant. 

As in ASTRID, most of the TNG systems lie in the active+star-forming quadrant at $z=2$, and the less massive central BHs along the $M_\mathrm{BH} - M_\star$ relation typically remain there as time progresses. The most massive BHs again form a horizontal band mostly in the inactive+quiescent quadrant. However, compared to ASTRID, this band is larger, extends more into the other quadrants, and is already populated by $z=2$. This is likely due to the lack of a redshift cutoff and the lower critical BH mass for AGN kinetic feedback, as discussed in Sec. \ref{sec:Mbh-Mstar}. Because of the absence of wandering BHs, there are significantly fewer systems in the inactive+star-forming quadrant. 

Fig. \ref{fig:Rbh_sBHAR-sSFR_tracks} emphasizes that most central BHs move uniformly from the active+star-forming quadrant toward the inactive+quenched quadrant. In contrast, wandering BHs fall into the inactive+star-forming quadrant, and some spread into the inactive+quenched quadrant. Their motion through the $\mathrm{sBHAR} - \mathrm{sSFR}$ plane is significantly less uniform, reflecting their changing hosts. 

\section{Comparison with the local empirical relation} \label{sec:observations}

\begin{figure}
    \centering
    \includegraphics[width=1\columnwidth]{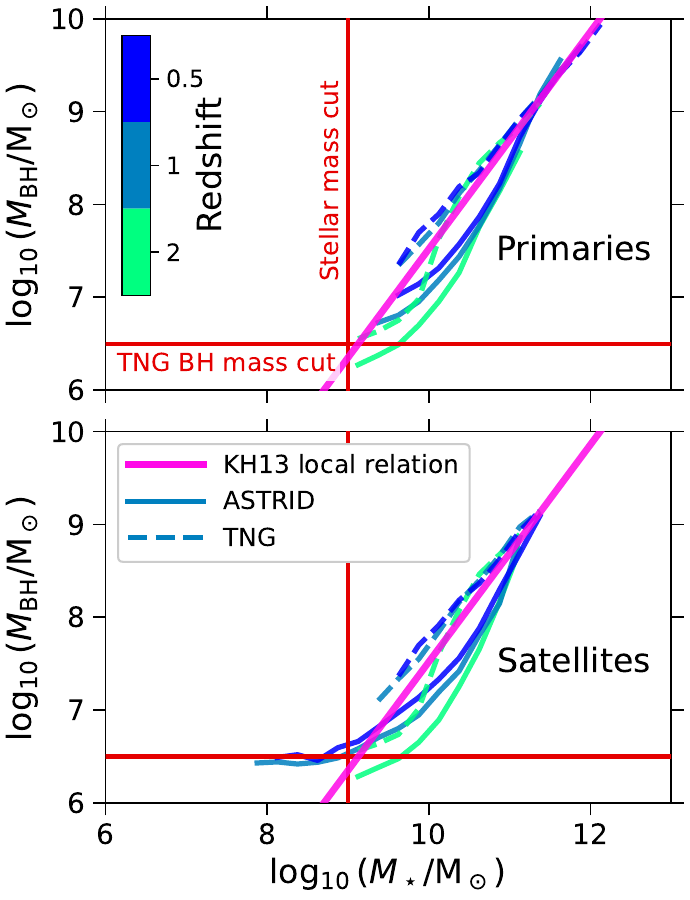}
    \caption{Median $M_\mathrm{BH} - M_\star$ relation for the central BHs in our sample, separated by simulation, redshift, and host galaxy type. The local ($z \sim 0$) empirical relation from KH13 is shown for comparison. Both simulations match KH13 reasonably well, especially at higher masses. TNG is closer at the low-mass end.}
    \label{fig:Median_Mbh-Mstar_lines}
\end{figure}

In Fig. \ref{fig:Median_Mbh-Mstar_lines}, we show the median $M_\mathrm{BH} - M_\star$ relation at $z=2,1,0.5$ for the central BHs in our sample, compared to the local empirical relation from KH13. As previewed in Sec. \ref{sec:Mbh-Mstar}, we only used BHs with galactocentric distances below $3 \, h^{-1} \, \mathrm{ckpc}$ (in ASTRID, this is twice the gravitational softening length, which is also the maximum separation for a BH merger). Medians were calculated in stellar mass bins of width $0.25 \, \mathrm{dex}$, and bins containing $\leq 10$ galaxies were excluded.

The median relation in the simulations shows minimal redshift evolution, and is relatively close to the empirical relation, consistent with our findings from Sec. \ref{sec:Mbh-Mstar}. In ASTRID, the slope is moderately steeper than the KH13 relation at intermediate masses and shallower at the low-mass end. The offset is slightly larger for higher redshifts.

We note that this work considers only galaxies that host BHs. A complementary diagnostic is the BH occupation fraction -- the fraction of galaxies that contain a BH, as a function of stellar mass. We investigate this topic in a subsequent paper \citep{Weller+2026}.

\section{Conclusions} \label{sec:conclusion}

\begin{figure}
   \centering
    \includegraphics[width=1\columnwidth]{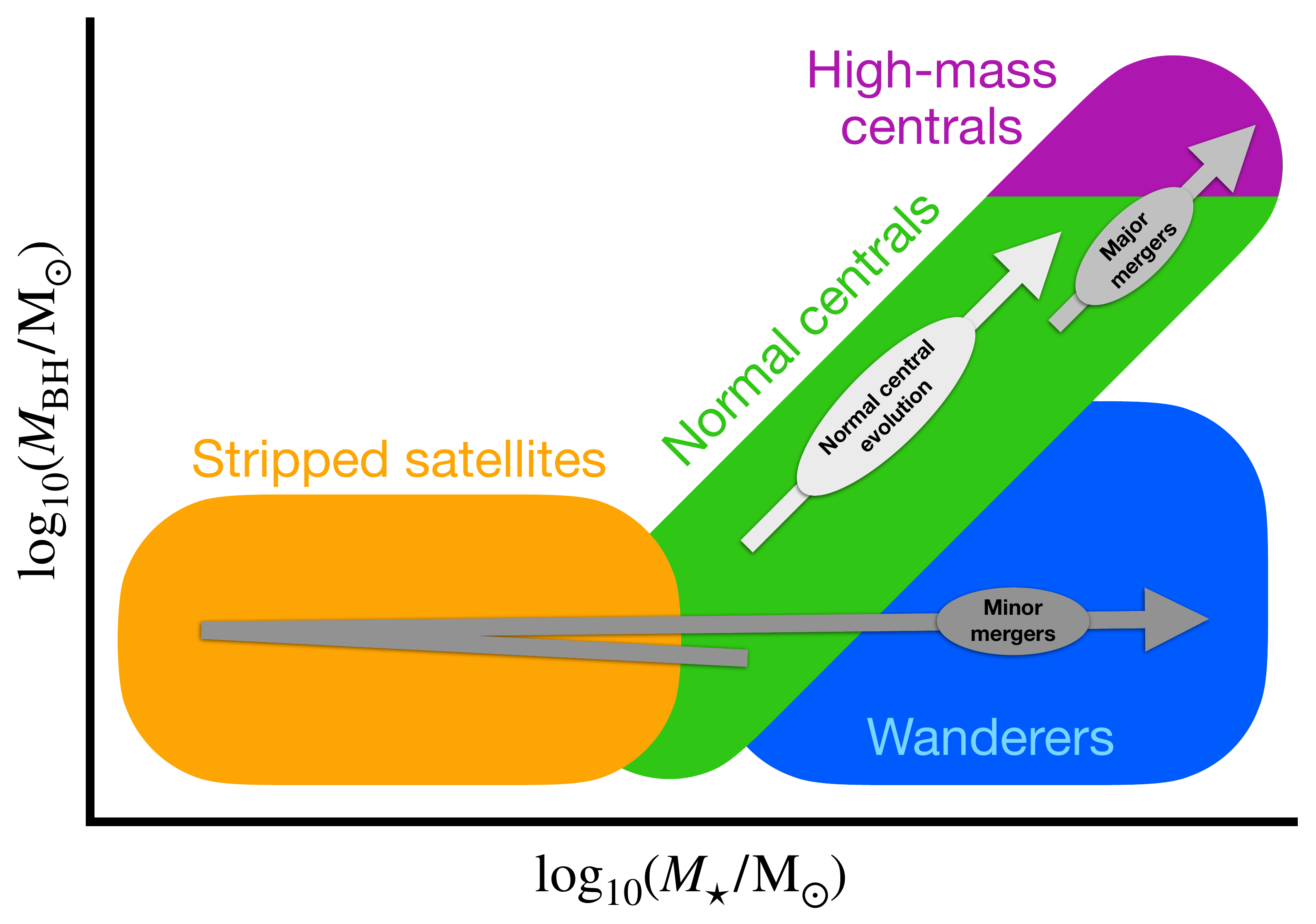}
    \caption{Schematic showing four regions of $M_\mathrm{BH} - M_\star$ space and the distinct evolutionary paths that build them, as summarized in Sec. \ref{sec:conclusion}. The stripped satellites and wanderers are not present in TNG due to its BH repositioning.}
    \label{fig:Schematic}
\end{figure}

In this work, we study the evolution of BHs and their host galaxies in the $M_\mathrm{BH} - M_\star$ and $\mathrm{sBHAR} - \mathrm{sSFR}$ planes from $z=2$ to $z=0.5$, using two cosmological simulations that adopt different prescriptions for BH dynamics and AGN feedback. In contrast to most previous studies, we follow the BH population rather than their host halos or galaxies, and trace individual BHs through cosmic time as they accrete, merge, and interact with their changing environments. In addition to central BHs in primary galaxies, we follow wanderers and BHs in satellites. This approach provides a more comprehensive picture of the evolutionary pathways of massive BH populations at late cosmic times when the overall gas reservoirs have started to deplete.

We summarize our key findings by interpreting four regions in $M_\mathrm{BH} - M_\star$ space, shown in Fig. \ref{fig:Schematic}.
\begin{itemize}
    \item \textbf{Normal centrals:} These BHs reside at the centers of primary and satellite galaxies. They grow upward along a linear $M_\mathrm{BH} - M_\star$ relation that, especially at higher masses, is already roughly consistent with local observational results by $z=2$. They are actively accreting and are hosted by star-forming galaxies, showing relatively low scatter in sBHAR and sSFR. Some of the BHs -- especially those in primary galaxies -- experience mergers, while others grow mostly via accretion.
    \item \textbf{High-mass centrals:} These are central BHs in primary galaxies, and they lie at the high-mass end of the linear $M_\mathrm{BH} - M_\star$ relation. Their growth is dominated by BH mergers that occur due to major galaxy mergers. AGN feedback suppresses BH accretion and star formation, resulting in low sBHAR and sSFR. The sBHAR values span a narrow range and are typically higher than those of the wanderers and stripped satellites.
    \item \textbf{Stripped satellites:} These satellite galaxies start on the low-mass end of the linear $M_\mathrm{BH} - M_\star$ relation but are tidally stripped by nearby galaxies during minor galaxy mergers. This shifts the galaxies toward lower $M_\star$, where they form a clump characterized by overmassive central BHs. Due to the loss of cold gas, stripped satellites typically have no star formation, and their central BHs have low accretion rates. Like most of the BHs in the lower-mass satellite galaxies, they do not experience significant BH merger-driven growth.
    \item \textbf{Wanderers:} These BHs originate along the linear $M_\mathrm{BH} - M_\star$ relation in satellites that later merge into more massive (usually primary) galaxies. As a result, they end up undermassive compared to the linear relation, forming a distinct clump that extends from the low-mass end of the relation toward larger $M_\star$. Due to their low-density environments, these BHs experience minimal accretion and few mergers. They cover a range of sSFR values (the sSFR is determined by the mass of the host and the AGN feedback from its central BH).
\end{itemize}

These regions demonstrate that massive BHs with different evolutionary histories populate distinct areas in $M_\mathrm{BH} - M_\star$ space. The first two regions exist in both ASTRID and TNG, though their extents appear to depend on the implementation of AGN kinetic feedback. In ASTRID, more detailed modeling of BH dynamics gives rise to wandering BHs and tidally stripped satellite galaxies that lie outside the typical $M_\mathrm{BH} - M_\star$ relation, while in TNG, these populations are absent due to BH repositioning.

We note that this work does not consider the evolution of dwarf galaxies and their BHs. Instead, to avoid resolution and seeding issues (see Sec. \ref{sec:selection}), we used BHs with masses above $10^6 \, \mathrm{M}_\odot$ (ASTRID) or $10^{6.5} \, \mathrm{M}_\odot$ (TNG) whose host galaxies reach $M_\star > 10^9 \, \mathrm{M}_\odot$ by $z=2$. As such, we cannot comment on truly low-mass galaxies, and all mentions of the ``low-mass end'' of the $M_\mathrm{BH} - M_\star$ relation in this work simply refer to the lowest masses considered in our study. 

Our results emphasize the role of galaxy interactions, BH dynamics, and AGN feedback in shaping the growth of massive BHs and driving departures from the linear $M_\mathrm{BH} - M_\star$ relation at late times. We show that a BH's position in the $M_\mathrm{BH} - M_\star$ and $\mathrm{sBHAR} - \mathrm{sSFR}$ planes can provide insight into its evolutionary history. Cosmological simulations with higher resolution that implement a range of more realistic ab initio seeding prescriptions will enable further investigation of the low-mass end of the $M_\mathrm{BH} - M_\star$ relation. 

\vskip -20pt 
\begin{acknowledgements}
We thank the anonymous referee for constructive comments that improved the paper. We thank Akaxia Cruz, Dylan Nelson, and Suvi Gezari for helpful discussions. E.J.W. acknowledges support from the National Science Foundation (NSF) Graduate Research Fellowship Program under Grant No. DGE-2139841. Any opinions, findings, conclusions, or recommendations expressed in this material are those of the authors and do not necessarily reflect the views of the NSF. P.N. acknowledges support from the Gordon and Betty Moore Foundation and the John Templeton Foundation, which fund the Black Hole Initiative (BHI) at Harvard University, where she is a PI. P.N. also acknowledges support from STScI/NASA via grant JWST-GO-03293024. C.J.B. is supported by an NSF Astronomy and Astrophysics Postdoctoral Fellowship under award AST-2303803. This research award is partially funded by a generous gift of Charles Simonyi to the NSF Division of Astronomical Sciences. The award is made in recognition of significant contributions to Rubin Observatory’s Legacy Survey of Space and Time. S.D. gratefully acknowledges NSF grant AST-2407063. 
\end{acknowledgements}

\software{Astropy \citep{astropy:2013, astropy:2018, astropy:2022}, Matplotlib \citep{Hunter_2007}, NumPy \citep{Harris+2020}, pandas \citep{McKinney2010, pandas_development_team_2024}, SciPy \citep{Virtanen+2020}}

\appendix

\section{Halo mass distributions} \label{app:halo_masses}

\begin{figure}
    \centering
    \includegraphics[width=1\columnwidth]{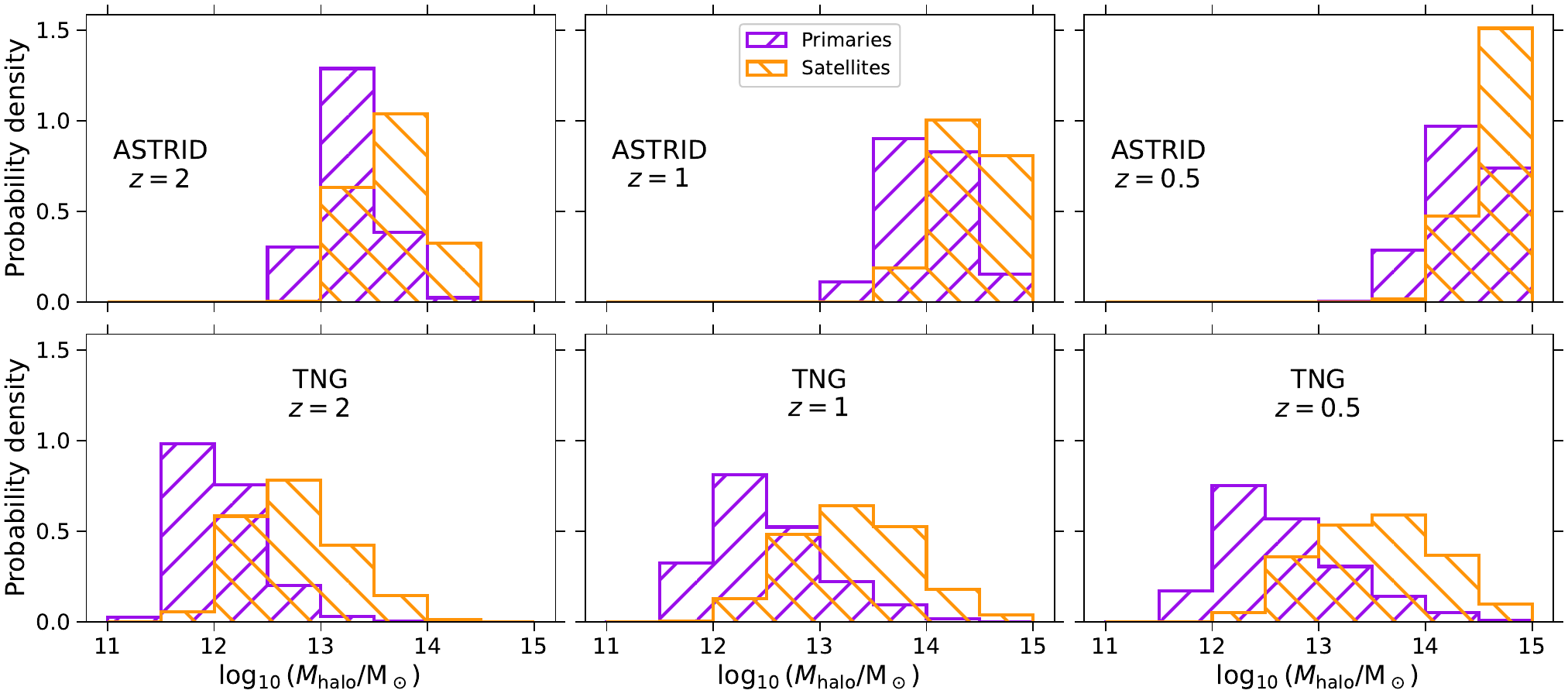}
    \caption{Mass distributions for the halos containing the host galaxies of our BH sample, separated by simulation, redshift, and host galaxy type. The halo masses are generally lower for primaries than for satellites, and lower in TNG than in ASTRID.}
    \label{fig:Mhalo_hist}
\end{figure}

In Fig. \ref{fig:Mhalo_hist}, we show the total masses of the halos containing our host galaxies. We see that the satellites tend to have more massive host halos than the primaries. This is because we set a fairly large minimum stellar mass of $10^9 \, \mathrm{M}_\odot$ at $z=2$, so only very massive halos would be expected to have satellite galaxies of this size. \textit{Primary} galaxies with $M_\star > 10^9 \, \mathrm{M}_\odot$ are relatively more common. The halo mass distributions are lower in TNG than in ASTRID. At a given halo mass, ASTRID and TNG halos have similar total stellar mass, but the ASTRID halos contain more galaxies, so the typical galaxy stellar mass (measured within twice the galaxy's stellar half-mass radius) is smaller. This explains why our host galaxies, selected to have $M_\star > 10^9 \, \mathrm{M}_\odot$ at $z=2$, can be found in smaller halos in TNG than in ASTRID.

\section{Supplemental $M_\mathrm{BH} - M_\star$ maps} \label{app:extra_plots}

\begin{figure}
    \centering
    \includegraphics[width=0.67\columnwidth]{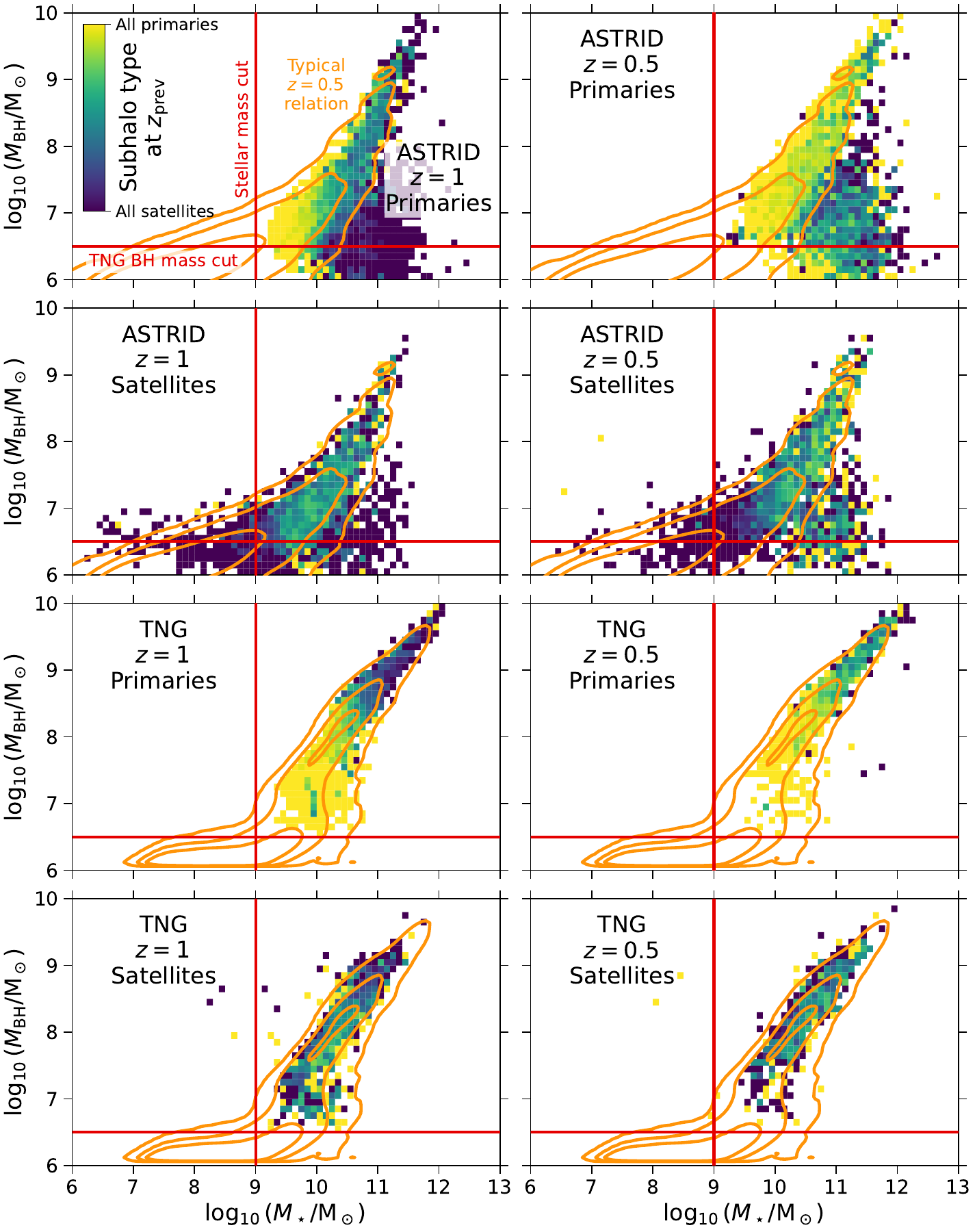}
    \caption{Binned $M_\mathrm{BH} - M_\star$ maps at $z=1$ and $z=0.5$, with each bin colored by the fraction of its BHs that were hosted by a primary galaxy at the previous snapshot ($z=2$ or $z=1$, respectively). Most stripped satellites have been satellites since $z = 2$, and most wanderers originated in satellites at $z=2$. The most massive central BHs in primaries tend to experience mergers with BHs from satellites.}
    \label{fig:Primary_Mbh-Mstar}
\end{figure}

\begin{figure}
    \centering
    \includegraphics[width=1\columnwidth]{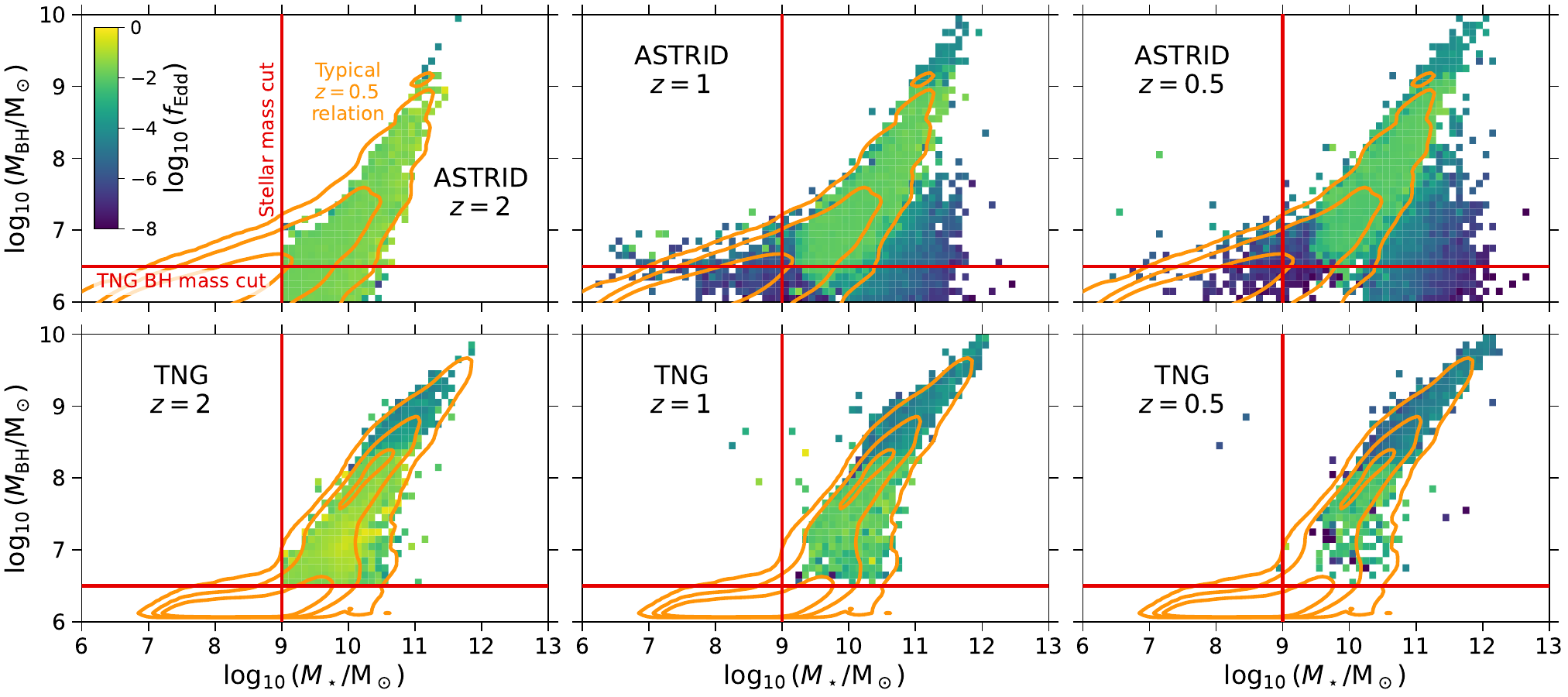}
    \caption{Binned $M_\mathrm{BH} - M_\star$ maps for our BH sample, with each bin colored by the median value of $\log_{10}{(f_\mathrm{Edd})}$, where $f_\mathrm{Edd} = \dot{M}_\mathrm{BH} / \dot{M}_\mathrm{Edd}$ is a BH's Eddington ratio. Panels are separated by simulation and redshift. The most massive galaxies and the stripped satellites have low BH accretion rates due to the removal of gas by AGN feedback and tidal stripping, respectively. The wandering BHs have minimal accretion due to their low-density environments. The actively accreting BHs have an approximately uniform Eddington ratio that decreases with time.}
    \label{fig:fEdd_Mbh-Mstar}
\end{figure}

Figs. \ref{fig:Primary_Mbh-Mstar} and \ref{fig:fEdd_Mbh-Mstar} are binned $M_\mathrm{BH} - M_\star$ maps like those shown in Figs. \ref{fig:Rbh_Mbh-Mstar}, \ref{fig:Merger-gain_Mbh-Mstar}, and \ref{fig:sSFR_Mbh-Mstar}. They provide further support for the findings discussed in Sec. \ref{sec:Mbh-Mstar}. In Fig. \ref{fig:Primary_Mbh-Mstar}, we color each bin by the fraction of BHs that were hosted by a primary galaxy at the previous snapshot. This is an alternative way of viewing the evolutionary paths illustrated in Fig. \ref{fig:Mbh-Mstar_tracks}. Fig. \ref{fig:fEdd_Mbh-Mstar} shows the median BH Eddington ratio in each bin, analogous to Fig. \ref{fig:sSFR_Mbh-Mstar}, which showed the sSFRs of the BH host galaxies.

\bibliography{sample701}{}
\bibliographystyle{aasjournalv7}

\end{document}